Article

# Extracting and Measuring Uncertain Biomedical Knowledge from Scientific Statements

Xin Guo [1,2,3,4 #], Yuming Chen [5,6 #], Jian Du [3,7*], Erdan Dong [1,2,3,4,8]

[1] Department of Cardiology and Institute of Vascular Medicine, Peking University Third Hospital, Beijing, China.
[2] NHC Key Laboratory of Cardiovascular Molecular Biology and Regulatory Peptides, Beijing, China.
[3] Key Laboratory of Molecular Cardiovascular Science, Ministry of Education, Beijing, China.
[4] Beijing Key Laboratory of Cardiovascular Receptors Research, Beijing, China.
[5] Department of Epidemiology and Biostatistics, School of Public Health, Peking University, Beijing, China
[6] Medical Informatics Center, Peking University, Beijing, China
[7] National Institute of Health Data Science, Peking University, Beijing, China
[8] Institute of Cardiovascular Sciences, Peking University, Beijing, China.

[*] Corresponding author, Jian Du, E-mail: dujian@bjmu.edu.cn
[#] Contributed equally

## Abstract

**Purpose:** There is an increasing need for computable biomedical knowledge since the information overload of scientific literature which is generally expressed in unstructured natural language. This study aims to develop a novel approach to extracting and measuring uncertain biomedical knowledge from scientific statements.

**Design/methodology/approach:** Taking cardiovascular research publications in China as a sample, we extracted the SPO triples as knowledge unit and the hedging/conflicting uncertainties as the knowledge context. We introduced Information Entropy and Uncertainty Rate as potential metrics to quantity the uncertainty of biomedical knowledge claims represented at different levels, such as the SPO triples (micro level), as well as the semantic type pairs (micro-level).

**Findings:** The results indicated that while the number of scientific publications and total SPO triples showed a liner growth, the novel SPO triples occurring per year remained stable. After examining the frequency of uncertain cue words in different part of scientific statements, we found hedging words tend to appear in conclusive and purposeful sentences, whereas conflicting terms often appear in background and act as the premise (e.g., unsettled scientific issues) of the work to be investigated.

**Research limitations:** Using cue words to represent textual uncertainty of biomedical knowledge may lead to a small amount of noise.

**Practical implications:** Our approach identified major uncertain knowledge areas, such as diagnostic biomarkers, genetic characteristics, and pharmacologic therapies surrounding cardiovascular diseases in China. These areas are suggested to be prioritized in which new hypotheses need to be verified, and disputes, conflicts, as well as contradictions to be settled further.

**Originality/value:** We provided a novel approach by combining natural language processing, computational linguistics with informetric methods to extracting and measuring uncertain knowledge from scientific statements.

*Keywords:* Uncertain Knowledge; Information Entropy; Natural Language Processing; Cardiovascular Diseases; China

## Introduction

According to whether be written down and directly used by others, scientific knowledge can be divided into explicit knowledge and tacit knowledge. In medical context, medical knowledge is considered as explicit knowledge due to the strict training standard and education system of medicine (especially modern medicine), which can be recorded and directly used by others (Wyatt & Scott, 2020). Therefore, medical literature and clinical guidelines produced by medical research have become the primary carriers of medical knowledge. However, published digital literature is generally expressed and stored in the unstructured natural language form, regularly in PDF or HTML files, and is complicated to comprehend and computed by machines. A vast number of knowledge assertions implied in medical literature have not been effectively managed and utilized. There is an increasing need for computable biomedical knowledge since the information overload of scientific literature.

Several efforts have been made to fill the gap. The current model of computable medical knowledge includes 1) the knowledge unit extracted from unstructured scientific text, in the form of subject-predicate-object triples (SPO triples) called predications (Kilicoglu, Rosemblat, Fiszman, & Shin, 2020), and 2) computable knowledge objects from structured data, such as disease prediction models, diagnostic rules, and structured clinical guidelines generated from big medical data (Flynn, Friedman, Boisvert, Landis-Lewis, & Lagoze, 2018; Friedman & Flynn, 2019). This article primarily focuses on the extraction of semantic triples from unstructured text, including knowledge assertions from medical literature, to reduce information overload. However, the study of extracting SPO triples as computable knowledge units from unstructured scientific text has been overwhelmingly focusing on scientific knowledge per se. Biomedical knowledge claims are often expressed in an uncertain way as hypotheses, speculations, or even controversial and contradictory opinions, rather than explicit facts. Considering the SPO triples would be conceivably extracted from hypothetical, speculative statements or even conflicting and contradictory assertions, the knowledge status (i.e., the uncertainty) serves as an integral and critical part of scientific knowledge that has been largely overlooked.

Uncertain scientific knowledge refers to the knowledge originating from hypothetical, speculative statements or even conflicting and contradictory assertions. It is critical to understanding the incremental and transformative development of scientific knowledge (Chen, Song, & Heo, 2018; Li, Peng, & Du, 2021; Small, 2020). For example, it is not rare in medical research or clinical practice to encounter "medical reversals," in which prior studies that claimed some therapeutic benefits were contradicted by subsequent research. Based on the analysis by (Herrera-perez et al., 2019), from more than 3,000 randomized controlled trials (RCTs) published in three leading medical journals, including *Journal of the American Medical Association*, *Lancet*, and *New England Journal of Medicine*, 396 medical reversals were identified, of which cardiovascular disease was the most common category (20%) that reversed. Therefore, there are increasing calls for physicians, patients, and the whole health care system to acknowledge and accept the concept of uncertain medical knowledge (Simpkin & Schwartzstein, 2016). These inspired us to investigate the uncertain medical knowledge of cardiovascular diseases since there are frequent medical reversals in this area. And in order to narrow the whole dataset, we focus on cardiovascular research publications in China as a sample.

**Related Work**

*Subject-Predicate-Object triples: a structured representation for knowledge claims*

Computable knowledge must be structured, or, in other words, machine-readable firstly. A simple computable representation of knowledge is the semantic triple, which consists of two concepts that are related to each other through specific predications (i.e., verbs), such as CAUSES and TREATS. For instance, drug knowledge can be expressed as triples (Elkin et al., 2011). Based on triples extracted from PubMed articles and compared with triples from the Food and Drug Administration (FDA) drug labeling documents, one can identify novel and unreported medical knowledge in the literature (Malec & Boyce, 2020).

There are well-established tools for semantic triples extraction from biomedical text. Among them, the SemRep tool and SemMedDB knowledge database developed by the Semantic Knowledge Representation program of the National Library of Medicine are typical representatives. SemRep, the abbreviation for Semantic Representation, is a rule-based natural language processing tool based on standardized medical concepts, concept types (e.g., drug, disease), and semantic relations between concepts (e.g., TREATS) in the Unified Medical Language System (UMLS). SemRep is used in extracting SPO triples from natural language texts. The latest version of UMLS contains about 3.8 million concepts, 127 concept types, and 54 semantic relations. The SemMedDB knowledge database stores triples and its supporting sentences extracted by SemRep from the titles and abstracts of PubMed documents (Kilicoglu, Shin, Fiszman, Rosemblat, & Rindflesch, 2012). It is released annually and continuously improved, including modifications of incorrect concepts and relations extracted by the SemRep tool. SemRep and SemMedDB support a variety of clinical decision-making and applications, including medical diagnosis, medication re-utilization, literature-based discovery, and hypothesis generation, to help improve health outcomes. The SemRep tool is currently being redesigned to improve its overall performance (Kilicoglu et al., 2020). SemRep and SemMedDB form a fundamental database that realizes the extraction and storage of large-scale knowledge units and supports secondary advancement. For example, the MELODI Presto system (http://melodipresto.mercies.ac.uk) developed by UK researchers provides queries based on Web pages for triples and their supporting sentences in SemMedDB (Elsworth & Gaunt, 2021).

## *Nanopublication: a richer semantic representation for knowledge claims*

In recent years, the progress in the field of biosemantics has provided solutions for the representation of medical knowledge objects. The nanopublication model is proposed by Professor Barend Mons and his team from Leiden University in the Netherlands (Groth, Gibson, & Velterop, 2010; Mons et al., 2011).

The basic structure of the nanopublication model includes a) assertion, that is, the scientific claims represented by subject-predicate-object triples; b) provenance, referring to the author, institution, time, and location that propose or create the factual material (e.g., data, figure); c) publication information, or the metadata of a nanopublication itself, including the creator, date of creation, and version of a nanopublication. The three components are indispensable to ensure the integrity of the information and effectively improve the possibility of reusing scientific research information. Expressed in RDF format, the contents of these components are machine-readable. Surveys in the biomedical area indicate that there are about $10^{14}$ such assertions, with an enormous increase every year. Eliminating duplicate findings still amounts to $10^{11}$ canonical assertions and further processing yields about $10^6$ so-called knowlets which can be seen as core concepts in this endless space of assertions related with sets of different findings (Mons, 2019). Finally, this highly reduced space of knowlets can be used to draw conclusions, for example, for proper health treatments.

There are currently several applications of nanopublication. First, researchers provide their scientific results as nanopublications (stored on the platform http://nanopublication.org/wordpress/). The second is to convert existing relational databases (such as DisGeNet, a gene-disease association knowledge database) in the form of nanopublications (Fabris, Kuhn, & Silvello, 2020). The Open Pharmaceutical Triple Store (Open PHACTS) (Williams et al., 2012), is a pharmaceutical semantic discovery platform for storing and computing pharmaceutical concept triples. However, the nanopublication model has not been widely utilized in clinical research and practice.

In order to promote the reuse of nanopublications and allocate the credit to the contributor, scholars recently proposed a nanocitation format that can cite individual nanopublication. It is, therefore, possible to design bibliometric indicators and conduct fine-grained analysis of knowledge units (Fabris, Kuhn, & Silvello, 2019).

Professor Barend Mons and his team, who proposed the nanopublication model, also suggested aggregating the common assertion that appears in all nanopublications into a cardinal assertion for presenting a basic knowledge unit to reduce redundancy (Mons, 2019). This process combines all directly linked concepts around a central concept as a knowlet. We think that knowlet, or the knowledge subgraph, can represent an independent knowledge unit in the knowledge graph. Compared with the rapid growth of support text related to a particular assertion, the growth of the knowledge subgraph is relatively slower. For instance, a large amount of text may only involve one piece of knowledge unit. The knowledge subgraph is an independent digital object, and the smallest knowledge unit can be discoverable, accessible, interoperable, and reusable.

In this article, by combining informetric methods with computational linguistics, we will extract knowledge units from scientific statements in cardiovascular research publications in China. In our previous work, we have put forward a framework for medical knowmetrics using the SPO triples as the knowledge unit and the uncertainty as the knowledge context. But, how to measure the uncertainty level for a given SPO triple or concept pairs is not well resolved. Therefore, inspired by the work by (Chen, 2020), we will introduce the concept of information entropy and uncertainty rate to measure the level of uncertainty for a given knowledge unit.

## Methods

Figure 1 demonstrated research framework for extracting and measuring computable biomedical knowledge of cardiovascular research in China.

### *Extracting Computable Biomedical Knowledge*

The process consists of 1) collecting related publications, 2) extracting semantic predications (SPO triples) from those publications, and 3) classifying semantic types and relations for the extracted SPO triples.

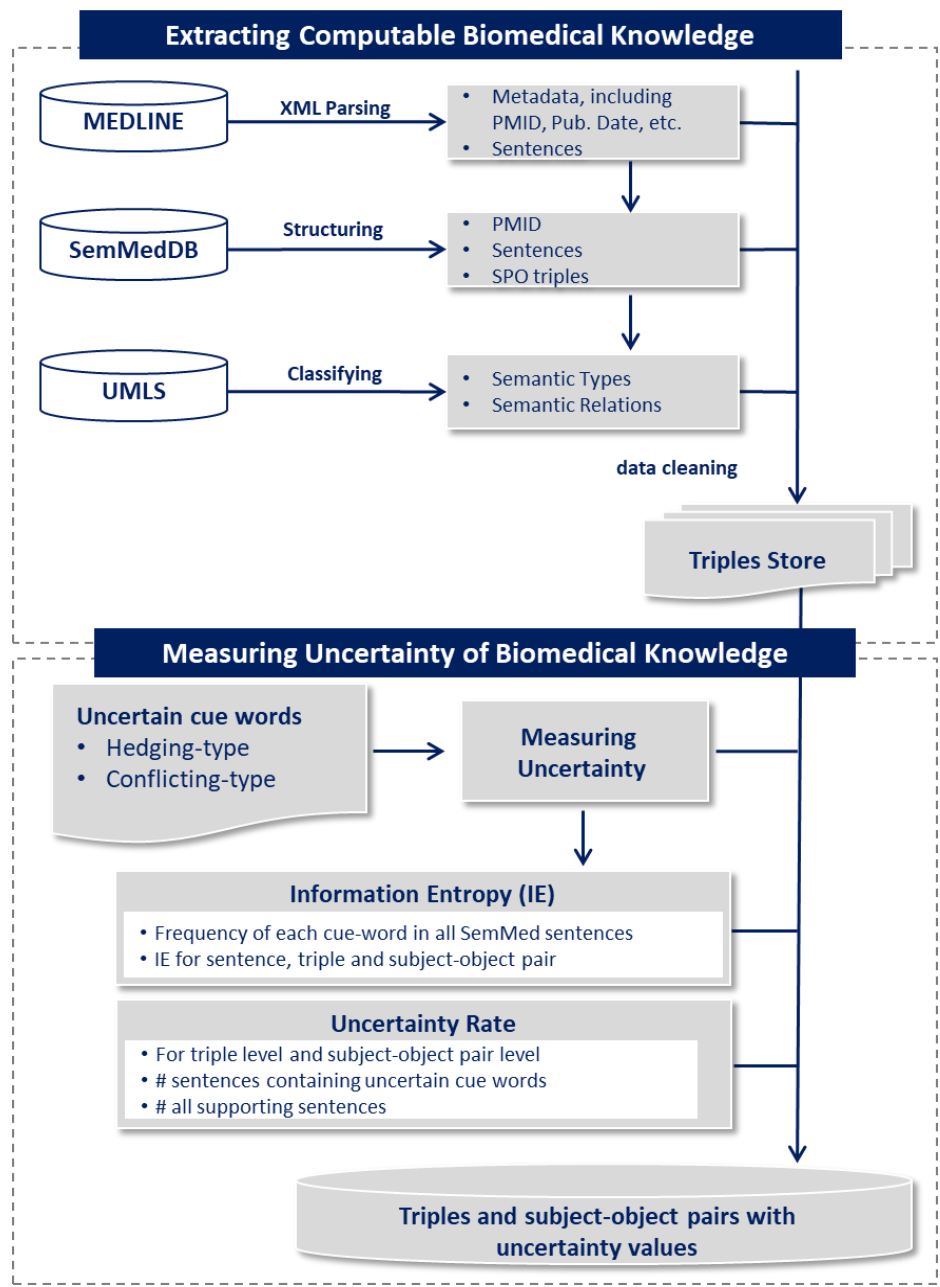

Figure 1. Research framework for extracting and measuring computable biomedical knowledge

*1) Collecting related publications*

We utilized the MEDLINE database to collect cardiovascular research publications during 2001-2020 in China. Cardiovascular diseases related Medical Subject Headings (MeSH) terms were combined to for archiving cardiovascular research publications. We then narrow our dataset by focusing on such research conducted in China or targeted on Chinese population. Cardiovascular research publications with 1) "China" or "Chinese" in the title or abstract and 2) Chinese language were included in our study. The acquisition for the metadata of included publications was performed on the 23rd Oct. 2020.

*2) Extracting semantic predications (SPO triples) from those publications*

To execute literature-derived computable biomedical knowledge, we first extracted semantic predications utilizing SemRep, an NIH program based on UMLS (Rindflesch

& Fiszman, 2003), and combined natural language processing method to handle the publications from MEDLINE. However, we observed multiple synonyms of a certain concept and several equivalent semantic triples from the same sentence in the extracting result. Considering the consistency and accuracy of biomedical knowledge extraction, we determined to retrieved sentences and the corresponding semantic predications from SemMedDB (Kilicoglu et al., 2012) by using PMID. The quality of semantic triples extracted by SemMedDB is significantly elevated compared with the result conducted by SemRep.

*3) Classifying semantic types and relations for the extracted SPO triples*

The UMLS Semantic Network, one of the UMLS Knowledge Sources, serves as an authority for the semantic types that are assigned to concepts in the Metathesaurus (see the UMLS Semantic Network: https://lhncbc.nlm.nih.gov/semanticnetwork/index.html). SemMedDB provides information about semantic types for each subject and object, as well as the semantic relations between them. Furthermore, the hierarchical information is included in the Semantic Network tree. In our work, semantic relations were grouped into eight major categories, i.e., "isa", "associated with", "physically related to", "spatially related to", "temporally related to", "functionally related to", "conceptually related to" and unclassified "others", such as "higher than", "converts to". And more than 130 semantic types were clustered into 20 third level groups, in order to make sure a general view for research evolution can be observed.

We compared the difference between the total and unique triples. The number of total triples is summarized by Predication ID. Unique triples are defined as those semantic predications with an identical subject, object, and predicate., among which the first occurrence time is labeled as the publish date. We utilized python programs for data processing and basic statistics analysis. In the established triples store, we included "functionally related to" semantic relations and additionally excluded "process of" type to leave the more informative triples and reduce the disturbance of meaningless triples.

## *Measuring Uncertainty of Biomedical Knowledge*

Uncertain biomedical knowledge was defined as triples extracted from sentences with uncertain cue words.

*1) Uncertain cue words*

The uncertain cue words list consists of hedging lexicon and conflicting lexicon. Supporting sentences of semantic triples are filtered with the hedging lexicon proposed by Bornmann L.'s article (Bornmann, Wray, & Haunschild, 2020) and Dakota's work of measuring disagreement in science (Murray et al., 2019) for conflicting lexicons. Hedging lexicon list includes "may/maybe", "possibl*", "potential", "seems", "perhaps", "likely" and "sometimes". The conflicting lexicon list contains "conflict*", "contradict*", "controvers*", "debat*", "disagree*", "disprov*", "no consensus",

"questionable*", "refut*", "uncertain" and "unknown", where "*" stands for wildcard character.

Semantic triples extracted from the title/abstract sentences could be understood as a highly generalized description on the content of article. Generally, the structured abstract of a given biomedical article includes Background, Objectives, Methods, Results and Conclusions, reflecting the process and elements of knowledge production. The information in Background usually is the premise and basics of the scientific issues to be investigated; Objectives tend to be unconfirmed hypotheses; Results convey evidence to support or attach the Objectives; and Conclusions are the knowledge claims, arguments and assertions arrived through the investigation. We determined different part of scientific statement by a machine identification check tags combining manual interpretation.

*2) Information Entropy (IE)*

Given a sentence, its uncertainty, *U(s)*, is defined as the information entropy (IE) regarding the probability *p(w)* of each of the uncertainty cue words, w. The probability for w is estimated based on the rate of its occurrences in the entire collection of sentences on SemMedDB. We have defined a set of hedging and conflicting cue words to represent two types of uncertainty. For a given SPO triple with multiple knowledge sources, we calculate its uncertainty, *U(t)* by computing the uncertainty of its supporting sentences.

$$U(s) = -\sum_{w \in s} p(w) \cdot \log(p(w))$$

$$U(t) = \sum_{1}^{n} U(s)$$

In SemMedDB, a triple can be linked to *n* sentences, $n \geq 1$. For such triple generated from more than two sentences, its uncertainty can be defined as the sum of the IE for all the associated sentences. The uncertainties of these sentences can be measured in terms of the presence of uncertainty cue words and how often these cue words appear in the entire collection of over 2,00 million sentences on the current version of SemMedDB. The information entropy measure gives different weight to each uncertain cue word according to the probability which each word appears.

*3) Uncertainty Rate*

Inspired by the concept of hedging rate (Small, 2018; Small, Boyack, & Klavans, 2019), in this article, uncertainty rate of a given triple or subject-object pair was defined as the ratio of the number of supporting sentences with uncertain cue words to the total number of supporting sentences. This measure gives equal weight to each uncertain cue word.

# Results

## *Overall results of SPO triples*

A total of 17,594 English articles and 17,196 Chinese articles published between 2001 and 2020 satisfied the specifications previously mentioned. Our final dataset included 29,800 articles and 313,837 unique sentences in total, excluding Chinese articles without English abstracts and articles with multiple PMIDs. In our dataset of articles, SemMedDB included 266,468 SPO triples extracted from the given range. Considering the redundancy of entirely equivalent SPO triples extracted from the same sentence and duplication of a specific article with distinct PMIDs, there were 259,067 SPO triples in our final dataset, in which 100,262 triples were unique. Since 2001, the semantic relation parsed from the article significantly enlarged, reaching 20,660 by the end of 2019. The number of total triples in 2001 increased by 6.1 times, novel triples (the new occurring triples each year) increased by 2.9 times. The number of publications itself also increased by 5.5 times during this period (Figure 2).

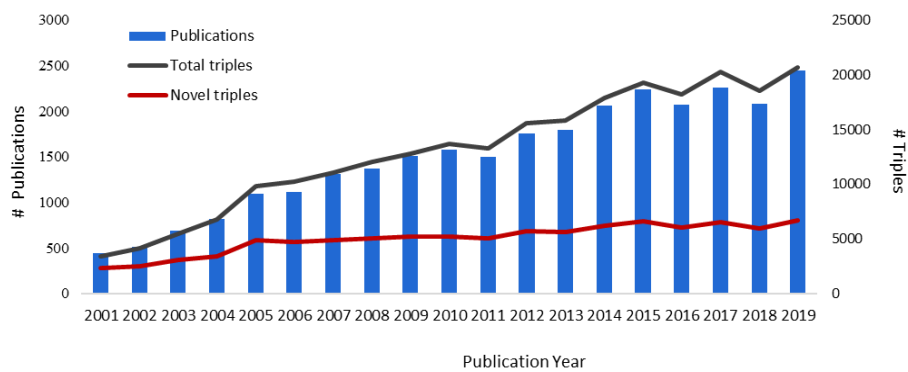

Figure 2. The number of publications, total triples, and novel triples

Semantic types and semantic relations were clustered to provide a general view on the evolution of cardiovascular research in China. Semantic relations were grouped into eight categories. Among them, functional relations accounted for the largest with a proportion up to 60% and a growing trend with time. In contract, the proportion of spatial relations decreased from 20% in 2001 to 15% in 2019. Other groups of relations were relatively small and remained stable over years (Figure 3).

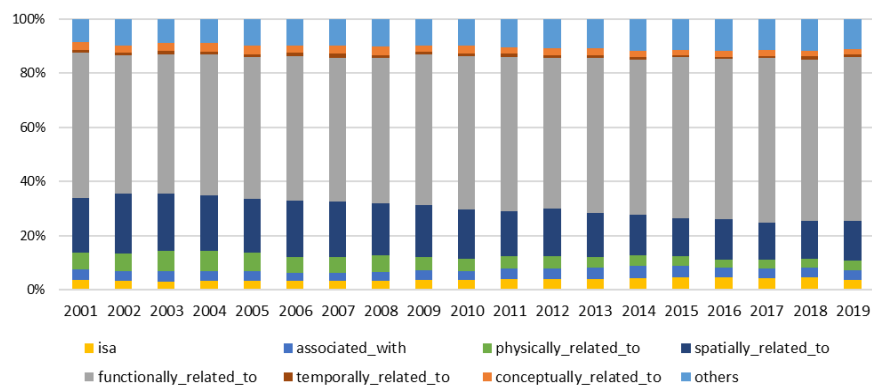

Figure 3. Distribution of different semantic relations linking the subject and object concepts

After checking the supporting sentences and corresponding SPO triples, we found that some triples were more likely to be general facts rather than scientific assertions. As showed in Example 1 and Example 2, "*Stem cells_Part_of _Marrow*" stands for "*MSCs (mesenchymal stem cells)*" and "*Hypertensive_ Process_of_Patients*" stands for "*hypertensive patients*". Example 3 is a representative of scientific assertion with meaningful and informative knowledge claims. In fact, they are functional related triples ("*NEG_predisposes*" is negative form of "*predisposes*"). So, in our final triple stores, semantic relations were narrowed to "*functionally_related_to*" by additionally excluding "*process_of*" in the follow-up data cleaning process, to improve theinfomativeness of extracted computable biomedical knowledge.

**Example 1**

SPO triple: "*Stem cells*"_Part_of _"*Marrow*"

Supporting sentence: *Subgroup analysis for each outcome measure was performed for the observing time point after the transplantation of MSCs.*

**Example 2**

SPO triple: "*Hypertensive*"_Process_of_"*Patients*"

Supporting sentence: *Our finding suggests that the CYP2C9*3 gene variant significantly alters the plasma concentration and acute DBP response at the 6-h point following irbesartan treatment in Chinese hypertensive patients.*

**Example 3**

SPO triples:

"*PGC gene, PGC*"_NEG_PREDISPOSES_"*Diabetes Mellitus, Non-Insulin-Dependent*"

"*PGC gene, PGC*"_NEG_PREDISPOSES_"*Hypertensive disease*"

Supporting sentence: *In conclusion, these results indicated that these two variations in the PGC-1alpha gene might not contribute to the risk of hypertension and type 2 diabetes in the Chinese population studied here.*

*Characterizing uncertain cue words*

We listed the uncertain cue words used in our study which consists of hedging lexicon and conflicting lexicon in Table 1. Supporting sentences with uncertain cue words were considered as knowledge sources with uncertain epistemic state. We utilized python programs for counting the appearing frequency for each cue word within all sentences in SemMedDB and calculating the value of IE.

Table 1. Uncertain cue words list of hedging and conflicting lexicons

| Cue word | Frequency in all SemMedDB sentences | IE |
|---|---|---|

| Hedging lexicon | | |
|---|---|---|
| may/maybe | 10286 | 0.00020693 |
| possibl* | 1751994 | 0.01703960 |
| potential | 2879336 | 0.02511068 |
| seems | 333677 | 0.00436449 |
| perhaps | 84058 | 0.00133387 |
| likely | 1052986 | 0.01132548 |
| sometimes | 119942 | 0.00181705 |
| Conflicting lexicon | | |
| conflict* | 175516 | 0.00252381 |
| contradict* | 46639 | 0.00079566 |
| controvers* | 208264 | 0.00292265 |
| debat* | 122332 | 0.00184838 |
| disagree* | 31384 | 0.00056055 |
| disprov* | 2517 | 0.00005780 |
| no consensus | 17907 | 0.00034016 |
| questionable* | 21159 | 0.00039480 |
| refut* | 9710 | 0.00019647 |
| uncertain | 227014 | 0.00314619 |
| unknown | 525536 | 0.00639116 |

We then analyzed the textual distribution in abstract of the two uncertain knowledge types. Combining structured abstract and knowledge generation process, we labeled 2711 sentences with 4 types: background, objectives, results, as well as conclusions. The frequency of uncertain cue words in different part of scientific statements (Figure 4) presented the different distribution of hedging knowledge and conflicting knowledge. Hedging lexicons tended to appear in conclusive and purposeful sentences, whereas conflicting lexicons tended to appear in background and act as premise of the investigated work.

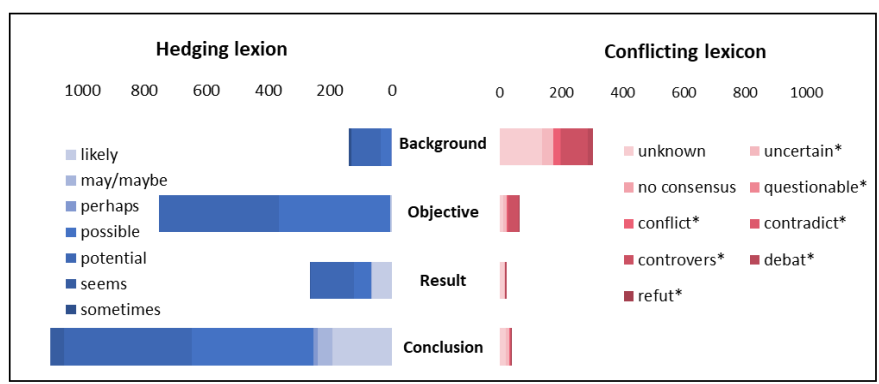

Figure 4. Distribution of hedging/Conflicting cue words in different parts of scientific statements

Examples of uncertain sentences and triples from different part of scientific statements are listed in Table 2. From the cognitive view, authors put forward cognitive judgment of existing knowledge in the background, presumption intention in objectives, evidences supporting the views of paper in results, and scientific claims and presupposition meaning in conclusions. It suggested that uncertain knowledge especially the conflicting knowledge usually was the premise and the start-point of scientific research, which can be considered as cognitive uncertainty. Moreover, author generally described object and conclusions with hedging lexicons, probably due to cautious attitude towards presupposition and modest attitude towards scientific claims.

Table 2. Examples of uncertain sentences and triples from different parts of scientific statements

| Statement Location | Supporting Sentences | SPO Triples |
|---|---|---|
| Premise (Background) | Liver X receptors (LXRs) play a central role in atherosclerosis; however, LXR activity of organic pollutants and associated potential risk of atherosclerosis have not yet been characterized. | liver X receptor _AFFECTS_ Atherosclerosis |
| Premise (Background) | Lipoprotein-associated phospholipase A2 (Lp-PLA2) is considered to be a risk factor for acute coronary syndrome (ACS), but this remains controversial. | Phospholipase A2 _PREDISPOSES_ Acute coronary syndrome |
| Hypothesis (Objective) | We therefore performed a case-control study investigating the possible relation between ACE gene polymorphisms and MVPS in Taiwan Chinese. | Mitral Valve Prolapse _ASSOCIATED_WITH_ gene polymorphism |
| Hypothesis (Objective) | Given the uncertainty regarding the relationship of C-reactive protein (CRP) and homocysteine (Hcy) to atherosclerotic burden, our aim was to determine whether CRP and Hcy are related to the presence of subclinical coronary plaque and stenosis. | Stenosis _ASSOCIATED_WITH_ C-reactive protein  Stenosis _ASSOCIATED_WITH_ homocysteine |
| Evidence (Results) | Ischemic heart disease was identified as the possible etiology of HF in a greater proportion of non-Chinese patients (47.7% vs. 35.3%; p < 0.001) whereas hypertension (26.1% vs. 16.1%; p < 0.001) and valvular heart disease (11.6% vs. 7.2%; p < 0.001) were relatively more common in Chinese patients. | Myocardial Ischemia _CAUSES_ Heart failure |

|  | Genetic polymorphisms of 4 genes, methylenetetrahydrofolate reductase (MTHFR) and apolipoprotein E (ApoE) have been demonstrated to associate with the increased risk for both MDD and stroke, while the association between identified polymorphisms in angiotensin converting enzyme (ACE) and serum paraoxonase (PON1) with depression is still under debate, for the existing studies are insufficient in sample size. | Peptidyl-Dipeptidase A _PREDISPOSES_ Cerebrovascular accident Peptidyl-Dipeptidase A _PREDISPOSES_ Major Depressive Disorder Arylesterase _PREDISPOSES_ Cerebrovascular accident Arylesterase _PREDISPOSES_ Major Depressive Disorder |
|---|---|---|
| Claims (Conclusions) | This study shows a significant association of hypertension susceptibility loci only in obese Chinese children, suggesting a likely influence of childhood obesity on the risk of hypertension. | Hypertensive disease _AFFECTS_ Obesity |
|  | Our data demonstrate that TrkB protects endothelial integrity during atherogenesis by promoting Ets1-mediated VE-cadherin expression and plays a previously unknown protective role in the development of CAD | ETS1 gene, ETS1 _INTERACTS_WITH_ cadherin 5 |

### *Measuring uncertainty by IE and Uncertainty Rate*

We compared two methods for measuring uncertainty. Uncertainty rate was defined as the ratio between the number of sentences containing uncertain cue words and the number of total sentences. Uncertainty rate could be measured at the triple level and at the subject-object pair level. Considering the different information content of cue words, IE considers different weights for each cue word based on frequency in all the sentences in SemMedDB. In our final dataset, 2612 SPO triples were extracted from 2711 supporting sentence. More 95% of triples were supported by only one uncertain related sentence. In other words, the value of IE for such SPO triple was determined by the uncertain cue words within the sentence. Besides, almost 87% of triples were supported by no more than 3 total sentences, indicating that uncertainty rate will concentrate on four values, i.e., 0/3, 1/3, 2/3, and 3/3. Given a small number of supporting sentences at the SPO triple level, we think both IE and uncertainty rate will not work well enough.

So, we turn to using IE and uncertainty rate to measure the uncertainty at the pairs of different semantic types level instead of the SPO triple level. More number of sentences will be aggregated if we only consider the semantic type co-occurrence pairs while do not take into account the specific predicate between the subject and object. In our dataset, 2196 SPO triples related to hedging lexicons were cluster into 407 semantic type co-occurrence pairs. As in Example 4, "*Brain-Derived Neurotrophic Factor_ PREDISPOSES _Diabetic Retinopathy*" was classified as knowledge claims describing the relationship between "Amino Acid, Peptide, or Protein" and "Disease or Syndrome". We listed the top 10 pairs with the highest IE (hedging) in Table 3, which also shows the value of IE, uncertainty rate and a corresponding SPO triple example.

**Example 4**

SPO triple: "*Brain-Derived Neurotrophic Factor*"_PREDISPOSES _"*Diabetic Retinopathy*"

Supporting sentence: *The present study demonstrated that decreased plasma levels of BDNF were independent markers for DR and VDTR in Chinese type 2 diabetic patients, suggesting a **possible** role of BDNF in the pathogenesis of DR complications.*

(IE: 0.042; Uncertainty rate: 0.667)

Table 3. Top 10 co-occurrence pairs of semantic types with the highest IE value (Hedging)

| Subject Type | Object Type | IE | Uncertainty Rate | SPO triple Example |
|---|---|---|---|---|
| Amino Acid, Peptide, or Protein | Disease or Syndrome | 2.93 | 0.71 | Brain-Derived Neurotrophic Factor_PREDISPOSES_Diabetic Retinopathy |
| Disease or Syndrome | Disease or Syndrome | 2.34 | 0.30 | Hypertensive disease_AFFECTS_Obesity |
| Therapeutic or Preventive Procedure | Disease or Syndrome | 2.13 | 0.19 | Acupuncture procedure_TREATS_Hypertensive disease |
| Gene or Genome | Disease or Syndrome | 1.79 | 0.59 | H19 gene, H19, HILS1_PREDISPOSES_Ischemic stroke |
| Pharmacologic Substance | Disease or Syndrome | 1.44 | 0.34 | Neuroprotective Agents_TREATS_Cerebrovascular accident |
| Disease or Syndrome | Pathologic Function | 0.99 | 0.50 | Myocardial bridging_PREDISPOSES_Thrombosis |
| Organic Chemical | Pathologic Function | 0.81 | 0.54 | Curcumin_PREVENTS_Inflammation |
| Health Care Activity | Disease or Syndrome | 0.75 | 0.28 | Intervention regimes_TREATS_Dementia, Vascular |
| Biologically Active Substance | Disease or Syndrome | 0.73 | 0.42 | Functional RNA_PREDISPOSES_Heart Diseases |
| Amino Acid, Peptide, or Protein | Pathologic Function | 0.73 | 0.87 | Adiponectin_PREVENTS_Myocardial Reperfusion Injury |

Similarly, 416 SPO triples related to conflicting lexicons were clustered into 128 semantic type co-occurrence pairs. The top 10 pairs with the highest IE (conflicting) were listed in Table 4. Combining two kinds of uncertainty, the highest uncertain knowledge areas were related to such pairs as:

- "Amino Acid, Peptide, or Protein"_"Disease or Syndrome",
- "Disease or Syndrome"_"Disease or Syndrome",
- "Therapeutic or Preventive Procedure"_"Disease or Syndrome",

- "Gene or Genome"_"Disease or Syndrome",
- "Pharmacologic Substance"_"Disease or Syndrome",
- "Disease or Syndrome"_"Pathologic Function", and
- "Biologically Active Substance"_"Disease or Syndrome", etc.

The above information indicated disease-centric uncertain knowledge since our sample is the cardiovascular research publications. We can learn that the diagnostic biomarkers, genetic characteristics, and pharmacologic therapies surrounding cardiovascular diseases in China remains at a higher uncertainty level in general. There are large number of new hypotheses in these areas that have not been verified, and there are also disputes and contradictions remain unresolved. These areas were the most active fields of cardiovascular research.

Table 4. Top 10 co-occurrence pairs of semantic types with the highest IE value (Conflicting)

| Subject Type | Object Type | IE | Uncertainty Rate | SPO Example |
|---|---|---|---|---|
| Disease or Syndrome | Disease or Syndrome | 0.16 | 0.64 | Ischemic stroke_COEXISTS_WITH_Primary Sjogren_s syndrome |
| Gene or Genome | Disease or Syndrome | 0.13 | 0.59 | ADIPOQ gene, ADIPOQ_ASSOCIATED_WITH_Ischemic stroke |
| Pharmacologic Substance | Disease or Syndrome | 0.13 | 0.88 | benidipine_TREATS_Hypertensive disease |
| Amino Acid, Peptide, or Protein | Disease or Syndrome | 0.12 | 0.50 | Receptors, C-Type Lectin_ASSOCIATED_WITH_Behcet Syndrome |
| Therapeutic or Preventive Procedure | Disease or Syndrome | 0.10 | 0.95 | Acupuncture procedure_TREATS_Cerebrovascular accident |
| Disease or Syndrome | Pathologic Function | 0.05 | 0.54 | Cerebrovascular accident_COEXISTS_WITH_Atrial Fibrillation |
| Finding | Disease or Syndrome | 0.05 | 0.88 | Impaired cognition_COEXISTS_WITH_Hypertensive disease |
| Pathologic Function | Disease or Syndrome | 0.05 | 0.53 | Pathogenesis_COEXISTS_WITH_Hypertensive disease |
| Disease or Syndrome | Genetic Function | 0.05 | 0.88 | Myocardial Infarction_ASSOCIATED_WITH_Polymorphism, Genetic |
| Biologically Active Substance | Disease or Syndrome | 0.03 | 0.53 | Triglycerides_ASSOCIATED_WITH_Atherosclerosis |

## Discussion and Conclusion

We have introduced the concept of a knowledge unit, defined as one identical SPO triple derived from multiple source sentences. SPO triple is a structured representation for knowledge assertions, less abundant than the unstructured text. We can quantity the increasing pattern of knowledge unit. In our dataset, while the number of scientific publications and total SPO triples shows a liner growth, the novel SPO triples occurring per year remain stable. In other words, new knowledge points or new research fields are not growing as fast as the number of scientific publications. This observation is in accordance with the results of "overall growth of science" within the whole publications in Web of Science. Whereas the number of publications grows exponentially, the conceptual territory of science, measured by the new appearing phases in the title expands only linearly (Fortunato et al., 2018).

We also added the contextual uncertainty information to the SPO triples since they were often extracted from hypothetical, speculative statements or even conflicting and contradictory assertions. Here we considered a transparent methodology that used a set of cue words to measure the level of contextual uncertainty rather than a machine learning approach. It is also consistent with a recent related work by (Kilicoglu, Rosemblat, & Rindflesch, 2017), which investigated the feasibility of assessing the factuality level of SemRep predications. They annotated semantic predications extracted from 500 PubMed abstracts with seven factuality values (fact, probable, possible, doubtful, counter-fact, uncommitted, and conditional). They proposed a rule-based, compositional approach that uses lexical and syntactic information to, and compared this approach to a supervised machine learning method that uses a rich feature set based on the annotated corpus. The results indicate that the compositional approach is more effective than the machine learning method in predicting factuality levels. This transparent approach allows us to easily scale our analysis to millions of scientific articles, while also being transparent and reproducible, interpretable of the results.

Two types of uncertainty, i.e., hedging, and conflicting, as well as a set of cue words were proposed for classifying the level of uncertainty. After examining the frequency of uncertain cue words in different part of scientific statements, we find hedging words tend to appear in conclusive and purposeful sentences, whereas conflicting words often appear in background and act as the premise (e.g., unsettled scientific issues) of the investigated work. Previous research has distinguished between two kinds of uncertainty: epistemic uncertainty about the past and present state of the world vs. uncertainty about the future that arises because we cannot know (van der Bles et al., 2019). It implies hedging lexicons associated with assumptions about the future, and conflicting lexicons associated with epistemic uncertainty. Distribution of hedging and conflicting lexicons in different part of scientific statements provides potential quantitative evidence of uncertain knowledge classification.

In order to detect where the uncertain knowledge exists, we built two measures to quantity the uncertainty level of SPO triples. Due to the small number of uncertain sentences in the triples we examined, we believe that both IE and uncertainty rate are not suitable for measuring the uncertainty of the triples with fewer sentences since there is not much research on these knowledge claims. As a result, we turn the measurement unit from triples to co-occurrence pairs of semantic types reflected by the subject and object concepts. A similar list of semantic type pairs was detected using both hedging and conflicting lexicons. Our approach identified major uncertain knowledge areas, such as diagnostic biomarkers, genetic characteristics, and pharmacologic therapies surrounding cardiovascular research in China. These areas need to invest more research to verify new scientific hypotheses, or to settle existing disputes, conflicts, and contradictions.

In conclusion, we provided a novel approach by combining natural language processing, computational linguistics with informetric methods to extracting and measuring uncertain knowledge from scientific statements. Both IE and uncertainty rate are potential metrics to quantity the uncertainty of biomedical knowledge claims represented at different levels, such as the SPO triples (micro level), as well as the semantic type pairs (micro-level).

Our study has some limitations. Uncertain cue words are applied to identify and measure the uncertainties in this article. In our current research, terms, such as "action potential" and "as soon as possible", do not indicate epistemic status of knowledge. We will optimize method and improve specificity of uncertain cue words. Uncertainty rate can help to identify more uncertain knowledge but with low resolution. High variation among the IE values of different cue words is unexplainable only based on its frequency in the whole set of biomedical sentences. We may first classify the clue words according to the degree of uncertainty, and then assign a probability. For example, the probability of the word "likely" is less than "very likely". Future research could refine the method for measuring uncertainty by creating metrics for the uncertain degree of cue word. We plan to apply our model to a broader range of data sets of medical knowledge, such as RCT registry information provided by clinicaltrials.gov, to further explore the feasibility of knowledge uncertainty computation using this model. And we plan to leverage the uncertainty-centric approaches to detecting research fronts, evaluating academic contributions, and improving the efficacy of computable knowledge-driven decision support.

## Conflict of interest

The authors declare that they have no conflict of interest.

## Acknowledgements

This work was funded by the National Natural Science Foundation of China (71603280,72074006, 82070235), the Beijing Municipal Natural Science Foundation (7191013), Peking University Health Science Center and the Young Elite Scientists Sponsorship Program by China Association for Science and Technology (2017QNRC001).

## Author contributions